\documentstyle[12pt]{article}

\newcommand{\be}{\begin{equation}}
\newcommand{\ee}{\end{equation}}
\begin{document}
\begin{titlepage}
\begin{flushright}
UWThPh-22-1998 \\
\today
\end{flushright}
\begin{center}
{\Large \bf A Treatment of the Schwinger Model\\
within Noncommutative Geometry}\\
{\small H. Grosse  \\
Institut for Theoretical Physics, University of Vienna, \\
Boltzmanngasse 5, A-1090 Vienna, Austria \\
P. Pre\v{s}najder\footnote{Supported by the "Fond zur F\"{o}rderung der
Wissenschaftlichen Forschung in \"{O}sterreich" project P11783-PHY
and by the Slovak Grant Agency VEGA project 1/4305/97.} \\
Department of Theoretical Physics, Comenius University \\
Mlynsk\'{a} dolina, SK-84215 Bratislava, Slovakia}\\[10pt]

\begin{abstract}
We describe a free spinor field on a noncommutative sphere starting
from a canonical realization of the algebra ${\cal U}(u(2|1))$ and a sequence
of $su(2)$-invariant embeddings of $su(2)$ representations. The gauge
extension of the model - the Schwinger model on a noncommutative sphere is
defined and the model is quantized. Due to the noncommutativity of the sphere,
the model contains only a finite number of modes, and consequently is
non-perturbatively UV-regular. The fermionic determinants and the effective
actions are calculated. The origin of chiral anomaly is clarified.
In the commutative limit standard formulas are recovered.
\end{abstract}
\end{center}
\end{titlepage}

\section{Introduction}

The basic notions of the noncommutative geometry were developed in
\cite{Con1}, \cite{Con2}, \cite{Con3}, and in the form of the matrix
geometry in \cite{DV}, \cite{DKM}. The essence of this approach consists in
reformulating first the geometry in terms of commutative algebras and
modules of smooth functions, and then generalizing them to their
noncommutative analogs. The notion of the space as a continuum of points
is lost, and this is expected to lead to an UV-regular quantum field
theory.

One of the simplest models for a noncommutative manifold is the
noncommutative (fuzzy) sphere. It was introduced by many authors using
various techniques, \cite{Ber},  \cite{Hoppe}, \cite{Mad}, \cite{GP1}. In
general, these are related to finite dimensional representations of the
$SU(2)$ group. Thus, the models in question are basically matrix models.
Quantum field theoretical models with a self-interacting scalar fields on
a truncated sphere were described in \cite{GKP1}, \cite{GKP2}. Since,
the fields posses only finite number of modes the models are UV-regular.

The basic ingredient of the noncommutative geometry developed in
\cite{Con1}, \cite{Con2} is the spectral triple $({\cal A}, D, {\cal H})$,
together with a chirality operator (grading $\Gamma$) and a charge
conjugation (antilinear isometry ${\cal J}$). Besides a noncommutative
algebra ${\cal A}$, the spectral consists of a Dirac operator $D$ acting
in the Hilbert space of spinor fields ${\cal H}$. The free spinor fields
on a noncommutative sphere in the spirit of noncommutative geometry were
introduced in \cite{GKP2} and \cite{GKP3} in the framework of
supersymmetric approach. The spectrum of the free Dirac operator was found
identical but truncated to the standard one on a commutative sphere.

Our aim is to demonstrate, that the axioms of the noncommutative geometry
can be implemented to a nontrivial field theoretical model. The model we
wish to describe is the noncommutative analog of the Schwinger model on
a sphere. The commutative quantum version was analyzed in detail in
\cite{Jay}. Its noncommutative matrix version was proposed in \cite{GM}, an
approach going behind matrix models was sketched in \cite{GKP4}. Recently
a systematic classical noncommutative formulation was found in \cite{K},
where a differential calculus on a supersphere in the commutative
and noncommutative cases is described in detail. Gauge models (for
topologically trivial field configurations) are described in both
commutative and noncommutative versions.

Here we shall first modify the supersymmetric descpription of
topologically nontrivial spinor field configurations on a noncommutative
sphere found in \cite{GKP2}. Instead of using in the formulation of the
model finite dimensional representations of the superalgebra $osp(2|1)$
(thus working in fact within the matrix model approach), we shall start
from an infinite dimensional canonical realization of the enveloping
algebra ${\cal U}(u(2|1))$. We introduce linear $su(2)$-invariant embeddings
of various sequences of finite dimensional representations of the even
$su(2)$ subalgebra (similar embeddings are important within a general
approach to the quantization of vector bundles proposed in \cite{Haw}; this
is natural since, our interpretation of fields on a nocommutative sphere is
technically close to the mentioned quantization). Only the evaluation of the
action is performed within particular finite dimensional representations. A
free Dirac operator is introduced and the model is gauged. After
specification of the gauge degrees of freedom the model is quantized. The
resulting model is well defined nonperturbatively and UV-regular. This
allows an exact analysis of various non-perturbative objects like
fermionic determinants and effective actions. The non-perturbative origin
of chiral symmetry  breaking is clarified. In the commutative limit the
standard formulas are reproduced (see e.g. \cite{Jay}).

The paper is organized as follows. In Section 2 we describe the free spinor
field on a noncommutative sphere within the $u(2|1)$ supersymmetric
formalism and we discuss linear embeddings of finite dimensional
representations of the even $su(2)$ subalgebra. In Section 3 we introduce
gauge degrees of freedom, and we present a field action for noncommutative
Schwinger model. In Section 4 we quantize nonperturbatively the
model, and then we calculate fermionic determinants and effective field
actions. Last Section 5 contains concluding remarks.

\section{Free spinor field}

First we summarize the commutative version of the model in question in the
$SU(2)$-invariant supersymmetric formulation (see \cite{Jay}
for $SU(2)$-invariant formalism, and \cite{GKP2}, \cite{GKP3}, \cite{K}
for its supersymmetric reformulation). In this approach the fields are
functions of two pairs $\chi_\alpha$, $\chi^*_\alpha$, $\alpha =1,2$, of
complex variables and of one anticommuting (Grassmannian) pair $a, a^*$:
\be
\Phi \ =\ \sum a^{\mu \nu}_{mn} \chi^{*m} \chi^n a^{*\mu} a^{\nu} \ ,\
\ee
where $n=(n_1 ,n_2)$ is a two component index and $\mu ,\nu = 0,1$ (we are
using a multiindex notations: $\chi^n = \chi^{n_1} ,\chi^{n_2}$,
$|n| = n_1 +n_2 $, $n! = n_1 !n_2 !$, etc.). The space $s{\cal H}_{k}$
of fields with a topological winding number $2k$ is defined as the space of
fields (1) with $2k=|n|-|m|+\mu -\nu$ fixed. Any field from $s{\cal H}_{k}$
can be expanded as
\be
\Phi \ =\ \Phi_0 (\chi ,\chi^*) \ +\ f(\chi ,\chi^*) a\ +\
g(\chi ,\chi^*) a^* \ +\ F(\chi ,\chi^*) \gamma \ ,
\ee
where $\gamma = a^* a - a a^*$, $\Phi_0 ,F \in {\cal H}_{k}$,
$f \in {\cal H}_{k+\frac{1}{2}}$ and $g \in {\cal H}_{k-\frac{1}{2}}$.
Here ${\cal H}_{k'}$ denotes the subspace of fields from $s{\cal H}_{k'}$
with $\mu =\nu = 0$.

The subspace of odd elements from $s{\cal H}_{k}$ is identified with the
space ${\cal S}_k$ of spinor fields with a given topological winding
number $2k$, i.e. any field from ${\cal S}_{k}$ can be expanded as
\be
\Psi \ =\ f(\chi ,\chi^*) a\ +\ g(\chi ,\chi^*) a^* \ ,
\ee
where $f \in {\cal H}_{k+\frac{1}{2}}$ and $g \in {\cal H}_{k-\frac{1}{2}}$.
The chirality operator $\Gamma$ in ${\cal S}_{k}$ is given as
$\Gamma =P_+ -P_-$, where $P_\pm $ are projectors
${\cal S }_k \to {\cal S }^{\pm}_k$ given by:
\be
P_+ \Psi \ =\ a\partial_a \Psi \ =\ fa \ ,\
P_- \Psi \ =\ a^*\partial_{a^*}  \Psi \ =\ ga^* \ .
\ee
The chirality operator takes in ${\cal S }^{\pm}_k$ the value $\pm 1$. The
charge conjugation ${\cal J}$ is defined as follows
\be
{\cal J}\Psi \ =\ {\cal J}(fa\ +\ ga^* )\ :=\ g^* a\ -\ f^* a^* \ =\
\Gamma \Psi^* \ =\ {\bar \Psi} \ .
\ee
Obviously, ${\cal J}:{\cal S}_{k} \to {\cal S}_{-k}$, and ${\cal J}^2 =-1$.
The inner product in ${\cal S }_k$ we define as
\be
(\Psi_1 , \Psi_2 )\ =\ \int d\mu {\bar \Psi}_1 \Psi_2 \ =\
\int d\nu (f^*_1 f_2 + g^*_1 g_2 )\ ,
\ee
where
$d\mu =\frac{1}{16\pi^2 r} d^2 \chi d^2 \chi^* da da^*
\delta (\chi^*_\alpha \chi_\alpha + a^* a - r)$ and
$d\nu =\frac{1}{2\pi r} d^3 x \delta (x^2_i - r^2 )$ are
normalized measures on a supersphere $sS^3$ and  sphere $S^2$ respectively.

In the superspace ${\bf C}^{2|1}$ we have a natural action of the
superalgebra $u(2|1)$ of matrices \cite{Ber2}
\be
\left( \begin{array}{cc}
X & 0 \\
0 & B \end{array} \right) \ +\ \frac{1-i}{\sqrt{2}}
\left( \begin{array}{cc}
0 & V \\
-V^* & 0 \end{array} \right) \ ,\ X\ =\ -X^* \ ,\ B \ =\ -B^* \ ,
\ee
written in a $(2|1)$-block diagonal form. In the space ${\bf C}^{2|1}$ a
super-Poisson bracket algebra can be introduced by postulating elementary
bracket relations among ${\bf C}^{2|1}$ coordinate functions
\be
[ \chi_\alpha ,\chi^*_\beta ] \ =\ \delta_{\alpha \beta} \ ,\
[ a, a^* ] \ =\ 1
\ee
(with all other elementary brackets vanishing). The $u(2|1)$ superalgebra
is then realized in terms of super-Poisson brackets by choosing its basis
in the following way:
\[
x_i \ =\ \frac{i}{2} \chi^+ \sigma_{i} \chi \ ,\
b\ =\ \chi^+ \chi \ +\ 2 a^* a \ ,\
\]
\be
v_\alpha \ =\ \chi_\alpha a^* \ ,\ \
{\bar v}_\alpha \ =\ \varepsilon_{\alpha \beta} \chi^*_\beta a\ ,\
r \ =\ \chi^+ \chi \ +\ a^* a \ .
\ee
The operators $x_i ,b$ are even generators and $v_\alpha ,{\bar v}_\alpha$
the odd ones of the superalgebra $su(2|1)$; $r$ is the central element
extending it to the $u(2|1)$ superalgebra. 

The $u(2|1)$ superalgebra is realized in the space of $s{\cal H}_k$ as the
adjoint Poisson-bracket superalgebra
$\{ X_i ,B,V_\alpha ,{\bar V}_\alpha ,R\}$:
\[
X_i \Psi \ =\ [ x_i ,\Psi ] \ ,\ B \Psi \ =\ [ b,\Psi ] \ ,\
R \Psi \ =\ [ r,\Psi ]\
\]
\be
V_\alpha \Psi \ =\ [  v_\alpha ,\Psi ] \ ,\
{\bar V}_\alpha \Psi \ =\ [ {\bar v}_\alpha ,\Psi ] \ .
\ee
Obviously, the space $s{\cal H}_k$ is an invariant space with respect
to (11). This action can be extended to the superalgebra
${\cal B}={\cal U}(u(2|1))={\cal B}^0 \oplus {\cal B}^1$ (the enveloping
algebra of the superalgebra ${\cal U}(u(2|1))$. Since the function
$r=\chi^+ \chi +\ a^* a$ is an invariant function with respect to this
action, all functions (fields) $\Phi$ can be factorized by the relation
$r = const$.

The spinor space ${\cal S}_k$ is invariant with respect to the action of
the even Poisson bracket subalgebra corresponding to ${\cal A}:={\cal B}^0$.
The free Dirac operator is an operator in ${\cal S}_k$ which has the
following form in terms of $u(2|1)$ generators
\be
D_0 \ =\ \frac {1}{2} \varepsilon_{\alpha \beta} (V_\alpha V_\beta \
+\ {\bar V}_\alpha {\bar V}_\beta )
\ =\ \varepsilon_{\alpha \beta} (\chi_{\beta} \partial_{\chi^*_\alpha}
a^* \partial_a \ +\ \chi^*_{\alpha}
\partial_{\chi_\beta} a \partial_{a^*} )\ .
\ee
We stress that this Dirac operator already contains a topological gauge
field term (the $k$-monopole magnetic field). Similarly, the chirality
operator can be expressed as
\be
\Gamma \ =\ B\ -\ R\ =\ a \partial_a \ -\ a^* \partial_{a^*} \ .
\ee
The chirality operator anticommutes with the Dirac operator:
$D_0 \Gamma =- \Gamma D_0$, i.e. the Dirac operator is a chiral odd operator
mapping ${\cal S}^{\pm}_k \to {\cal S}^{\mp}_k$. The free spinor field
action is defined by
\be
S[\Psi ,\Psi^* ]\ =\ \int d\mu {\bar \Psi} D_0 \Psi \ .
\ee

Now we shall describe a noncommutative version of the free spinor model. We
shall go beyond matrix models starting from the infinite dimensional
superalgebra ${\cal B} = {\cal U}(u(2|1)) ={\cal B}^0 \oplus {\cal B}^1$
and its even subalgebra ${\cal A}:={\cal B}^0$. Only the evaluation of
the action will be performed at some finite level.

We shall work within the canonical $\pi {\cal B}$ realization of the
superalgebra ${\cal B}$: we insert graded commuting variables $a, a^*$,
$\chi_{\alpha}$, $\chi^*_{\alpha}$, $\alpha = 1,2$ by annihilation and
creation operators satisfying graded commutation relations
\be
[ \chi_\alpha ,\chi^*_\beta ]\ =\ \delta_{\alpha ,\beta} \ ,\
[ a, a^* ] \ =\ 1
\ee
(all other elementary commutators vanish). They act in the Fock
(super)\-space
\be
s{\cal F}\ =\ \{ |n,\nu \rangle \ =\
\frac{1}{\sqrt{n!}} \chi^{*n} a^{*\nu} |0 \rangle \} \ =\
s{\cal F}^0 \ \oplus s{\cal F}^1 \ ,
\ee
possesses a natural grading with respect to the fermion occupation number
$\nu$ (in eq. (16) $n$ is a two-component multiindex, $\nu = 0,1$, and
$|0 \rangle $ is the corresponding normalized vacuum state:
$\chi_\alpha |0 \rangle = a|0 \rangle = 0$). In what follows we shall
use the notation ${\cal F} =s{\cal F}^0$, and we shall simply write
${\cal B}$ and ${\cal A}$ instead of $\pi {\cal B}$ and $\pi {\cal A}$.

The generators satisfying in $s{\cal F}$ the ${\cal U}(u(2|1))$ graded
commutation relations (10) are given by eqs. (9) (commuting parameters
are replaceded by annihilation and creation operators and graded Poisson
brackets by corresponding graded commutators). The Fock subspace
\be
s{\cal F}_N \ =\ \{ |n,\nu \rangle \ ,\ |n| + \nu = N \} \ =\
s{\cal F}^0_N \ \oplus s{\cal F}^1_N \ ,
\ee
is the carrier space of the unitary irreducible representation $\pi'_N$ of
$u(2|1)$ superalgebra. We shall use the notation ${\cal F}_N =s{\cal F}^0_N$.

We define the space $s{\cal H}_{k}$ of superfields with the winding number
$2k$ as the set of operators in the Fock space of the form (1) with
$2k=|n|-|m|+\mu -\nu$ fixed. Obviously, $\Phi \in s{\cal H}_{k}$ maps
any space $s{\cal F}_N$ into the space $s{\cal F}_M$, $M=N+2k$. The $u(2|1)$
superalgebra is realized in $s{\cal H}_{k}$ as the adjoint graded
commutator superalgebra (10). Any superfield from $s{\cal H}_{k}$ can be
expressed as in (2) with coefficients from ${\cal H}_k$ (however now
expressed in terms of anihilation and creation operators).

By restricting the action of superfields $\Phi \in s{\cal H}_k$ from
$s{\cal F}$ to $s{\cal F}_N$ we obtain a space of mappings
$s{\cal F}_N \to s{\cal F}_M$, $M=N+2k$. We shall denote them
by $s{\cal H}^{J'}_k$, $k=\frac{1}{2}(M-N)$,
$J'=\frac{1}{2}(M+N-1)=J-\frac{1}{2}$ (these relations among $M, N, J, J'$
and $k$ are assumed in what follows). From the definition of
$s{\cal H}^{J'}_k$ it follows that on $s{\cal H}^{J'}_k$ acts the direct
product ${\bar \pi}'_M \otimes \pi'_N$ of the irreducible representations
${\bar \pi}_M$ and $\pi_N$  of the superalgebra $u(2|1)$.

{\it Note}: Representations $\pi'_N$ of $u(2|1)$ are well known. They
corresponds to the atypical $(q,-q)$ representations of $su(2|1)$, in the
notation of \cite{Rit}: quadratic and cubic Casimir operators vanish and
they are classified by the value $R = N$ of the central element in $u(2|1)$
superalgebra. However, their products are indecomposable and have
quite pathalogical properties, \cite{Rit}. This causes problems with the
identification of auxiliary fields in the noncommutative case.

Such problems do not occur for the coefficient spaces ${\cal H}_{k}$ of
operators acting in ${\cal F}_N := s{\cal F}^{0}_N$. The subspaces
${\cal H}^J_k$ are defined as the spaces of mappings
${\cal F}_N \to {\cal F}_M$. In ${\cal H}^J_k$ we introduce the scalar
product
\be
(\Phi_1 ,\Phi_2 )^J_k \ =\ \frac{1}{J+1} {\rm Tr} (\Phi^*_1 \Phi_2 )\ .
\ee
In ${\cal H}^J_k$ acts the direct product representation
${\bar \pi}_M \otimes \pi_N$ of the irreducible $SU(2)$ representations
${\bar \pi}_N$ and $\pi_N$ in ${\cal F}_M$ and ${\cal F}_N$ respectively.
The direct product ${\bar \pi}_M \otimes \pi_N$ possesses a decomposition
to irreducible $SU(2)$ representations, and consequently
\be
{\cal H}^J_k \ =\ \bigoplus^J_{j=|k|} V^{Jj}_k \ ,
\ee
where $V^{Jj}_k$ denotes a $SU(2)$ representation space corresponding to a
spin $j$.

For each $J$ and $k$ we constructed in Appendix A an orthormal basis
$\{ D^j_{km}$, $j=0,1,\dots ,|m|\leq j \}$ in ${\cal H}_k$. The orthonormal
basis $\{ D^{Jj}_{km} = c^J_{kj} D^j_{km}$, $j=0,\dots ,J$, $|m|\leq j\}$
in ${\cal H}^J_k$ with respect to the scalar product (17) is determined
by rescaling coefficients $c^J_{kj}$ determined in Appendix A.

We define the isometrical projection operator
$C^J_k :{\cal H}_k \to {\cal H}^J_k$
by $C^J_k D^j_{km} =D^{Jj}_{km}$, and by
${C'}^J_k :{\cal H}^J_k \to {\cal H}_k$ we denote the reversed isometrical
embedding. We can extend this construction by taking $K \geq J$ and
defining the isometrical projections
$C^{KJ}_k :{\cal H}^K_k \to {\cal H}^J_k$, together with the reversed
isometrical embeddings
${C'}^{JK}_k :{\cal H}^J_k \hookrightarrow {\cal H}^K_k$. It holds
\be
C^{KJ}_k \ =\ C^J_k {C'}^K_k \ ,\ {C'}^{JK}_k \ =\ C^K_k {C'}^J_k \ .
\ee
These are the key objects of our approach which give the prescription
how  are related to each other for various values of $J$ the (matrix)
realizations in ${\cal H}^J_k$ of a given field from ${\cal H}_k$.

Similarly as in the graded commutative case, we define the space
${\cal S }_k$ of spinor fields with a topological winding number $2k$ as
the odd subspace of the space $s{\cal H}_k$. Such spinor fields can be
expanded as
\be
\Psi \ =\ f(\chi ,\chi^* ) a\ +\ g(\chi ,\chi^* ) a^* \ ,
\ee
where $f\in {\cal H}_{k+\frac{1}{2}}$ and $g\in {\cal H}_{k-\frac{1}{2}}$.
The chirality operator $\Gamma :{\cal S }_k \to {\cal S }_k$ and the charge
conjugation ${\cal J}:{\cal S }_k \to {\cal S }_{-k}$ are defined in the
same way as in the commutative case.

The subspaces ${\cal S }^{J'}_k$ are again given by the restriction:
they are odd mappings from ${\cal H }^{J'}_k$, i.e. odd mappings
$s{\cal F}_N \to s{\cal F}_M$ with $M=J'+k+\frac{1}{2}$,
$N=J'-k+\frac{1}{2}$. In the space ${\cal S }^J_k$ we choose the inner
product
\be
(\Psi_1 ,\Psi_2 )^{J'}_k \ =\
\frac{1}{2J'+1} s{\rm Tr}_N ({\bar \Psi}_1 \Psi_2 ) \ ,
\ee
where $s{\rm Tr}_N$ denotes the supertrace in the space of mappings
$s{\cal F}_N \to s{\cal F}_N$.

Now we shall concentrate our attention to the space of mappings
${\cal S }_k \to {\cal S }_k$. Such mappings have the form
\be
\sum A_i B^o_i \Psi \ :=\ \sum A_i \Psi B^*_i \ ,
\ee
where $A_i$,$B_i$ are elements from $s{\cal H}_0$ such that all products
$A_i B_i$ are even ($B^o$ means the right multiplication of $\Psi$
by $B^*$). In the space of mappings ${\cal S }_k \to {\cal S }_k$ we have
a natural grading induced by the chirality operator: the mapping
$\sum A_i B^o_i$ is even (odd) if all  $A_i$ and $B_i$ are even (odd).
Obviously, ${\cal S }^{\pm}_k \to {\cal S }^{\pm}_k$ for the even mappings
and ${\cal S }^{\pm}_k \to {\cal S }^{\mp}_k$ for the odd ones.

In the noncommutative case the free Dirac operator is given by the first
equation in (11) but with graded Poisson brackets inserted by graded
commutators. The resulting expression can be rewritten in the form
\be
D_0 \ =\ v_\alpha {\bar v}^o_\alpha \ -\ {\bar v}_\alpha v^o_\alpha \ ,
\ee
which was proposed earlier in \cite{GKP2}. This is the simplest odd
mapping commuting with the $su(2)$ generators $X_i = x_i - x^o_i$,
$i = 1,2,3$.

The spectrum of the free Dirac operator contains:

(i) non-zero modes
\be
E^{J'\pm}_{jk} =\pm \sqrt{(j+k+\frac{1}{2})(j-k+\frac{1}{2})} \ ,\
j=|k|+\frac{1}{2},|k|+\frac{3}{2},\dots , J \ ,
\ee

(ii) and $2|k|$ zero-modes corresponding to $j=|k|-\frac{1}{2}$ (if
$k \neq 0$).\\
The corresponding eigenfuctions $\{ \Psi^{J'j\pm}_{km} \}$,
$j=|k|-\frac{1}{2},|k|+\frac{1}{2},\dots ,J'$, $|m| \le j$ are presented
in Appendix A. They form an orthonormal basis in ${\cal S}^{J'}_k$. Any
spinor field from ${\cal S}^{J'}_k$ can be expanded as
\be
\Psi \ =\ \sum_{j=|k|+1/2}^{J'} \sum_{|m|<j}
(a^{j+}_{km} \Psi^{J'j+}_{km} \ +\ a^{j-}_{km} \Psi^{J'j-}_{km} ) \ .
\ee

For a given $k$ and $J=J' +\frac{1}{2}$ the free spinor field action we
take as
\be
S^J_k [\Psi ,\Psi^* ] \ =\ (\Psi ,D_0 \Psi )^J_k \ .
\ee
                                                                   
The spinor field space	${\cal S }_k$ is obviously an
${\cal A}$-bimodule. The algebra ${\cal A}\otimes {\cal A}^o$
acts in ${\cal S }_k$ as left-multiplication by $A \in {\cal A}$ and
right-multiplication by $B^* \in {\cal A}$ (see (22)). The Dirac operator
acts in ${\cal A}\otimes {\cal A}^o$ as a commutator:
\be
D_0 (A B^o \Psi ) \ =\ [D_0 , A B^o ]\Psi \ +\ A B^o D_0 \Psi \ .
\ee
Thus, $D_0 : A B^o \to [D_0 , A B^o ]$. Consequently, the graded Leibniz
rule in ${\cal A}\otimes {\cal A}^o$ is satisfied. Obviously, this changes
the grading of the mapping. The charge conjugation ${\cal J}$ induces in
the space ${\cal S }_k \to {\cal S }_k$ an automorphism
$AB^o \to {\cal J}^{-1} AB^o {\cal J}$. It is easy to see that the Dirac
operator corresponds to an ${\cal J}$-odd mapping:
${\cal J}^{-1} D_0 {\cal J} = - D_0$.

\section{Gauge symmetry}

In the commutative case the gauge transformations of the spinor field are
given in the following way
\be
\Psi \ \to \ \omega \Psi \ ,\ \Psi^* \ \to \ \omega^* \Psi^* \ ,
\ee
where $\omega$ is an arbitrary unitary element from ${\cal H}_0$. To
guarantee the gauge invariance one should replace the free Dirac operator
$D_0$ by the full Dirac operator
\be
D\ =\ D_0 \ +\	q_o A \ .
\ee
Here $q_o$ is a dimensionless coupling constant related to the usual
coupling constant $q$ and the radius $r$ of the sphere in question as
$q_o = qr$. The compensating gauge field $A$ transforms inhomogenously
under gauge transformations
\be
A\ \to \ A\ +\ \omega [D_0 ,\omega^* ]\ .
\ee
Since, $D_0$ already contains the topological (k-monopole) gauge field, the
field $A$ is a globally defined gauge field. Such gauge fields can be
expressed in terms of the real prepotentials $\lambda \in {\cal H}_0$
corresponding to the pure gauge field $A$ (exact 1-form) and
$\sigma \in {\cal H}_0$ corresponding to the dynamical gauge field $A$
(coexact 1-form). The full Dirac operator can be written in terms of
${\tilde \Omega}= U Q P_+ + U Q^{-1} P_-$ with $U =\exp(i\lambda )$ and
$Q = \exp (q_o \sigma )$. The formula for $D$ can be rewritten as
\be
D\ =\ {\tilde \Omega} D_0 {\tilde \Omega}^* \ .
\ee

The square of the Dirac operator has the form
\be
D^2 \ =\ [\Delta (\Omega ) + q_o F(\Omega )] P_+ \ +\
[\Delta (\Omega ) - q_o F(\Omega )] P_- \ ,
\ee
where $\Delta (\Omega )$ is the Laplace operator on a sphere with the
gauge field included, and the field strenght operator $F(\Omega )$ is
\be
F(\Omega ) \ =\ q^{-1}_o (\chi^*_\alpha \partial_{\chi^*_\alpha} -
\chi_\alpha \partial_{\chi_\alpha} )\ +\ \Delta \sigma \ .
\ee
The differential term takes in ${\cal S}_k$ the constant value $q^{-1}_ok$
and it corresponds to the contribution of the $k$-monopole, the last term
is the field strenght generated by the dynamical field $\sigma$.

The pure gauge prepotential $\lambda$ can be gauged away, and there are no
restrictions on it. There are limitations on the dynamical prepotential,
since $\sigma$ enters the Schwinger model field action
\be
S[\Psi ,\Psi^* ,\Omega ]\ =\ \frac{1}{4} \int d\nu F^2 (\Omega ) \ +\
\int d\mu {\bar \Psi} D \Psi \ ,
\ee
which is obviously gauge invariant. Expanding $\sigma$ into spherical
functions (denoted as $D^j_{0m}$):
\be
\sigma \ =\ \sum_{j=0}^{\infty} \sum_{|m|\leq j} b^j_m D^j_{0m} \ ,
\ b^j_{-m} = (-1)^m b^{j*}_m \ ,
\ee
we obtain a contribution to the action proportional to
\[
\sum_{j=0}^{\infty} \sum_{|m|\leq j} j^2 (j+1)^2 |b^j_m |^2 \ .
\]
This is finite provided that
\be
b^j_m \ =\ o(j^{-5/2} )\ .
\ee

In the noncommutative case we identify the group of gauge
transformations with the group of unitary operators
$\omega \in {\cal H}_0$, and we postulate that the spinor fields
$\Psi \in {\cal S }_k$ transform under gauge transformations as follows
\be
\Psi \ \to \ {\tilde \omega} \Psi \ =\ \omega \omega^o \Psi \ =\
\omega \Psi \omega^* \ .
\ee
This is an unitary ${\cal J}$-invariant mapping in ${\cal S }_k$. In
order to obtain a gauge invariant action we introduce compensating
gauge degrees of freedom. We shall describe them by the operators of the
form $\Omega = UQ^{(+)} P_+  +UQ^{(-)} P_- $, where $U \in {\cal H}_0$ is
an unitary operator describing pure gauge degrees of freedom and
$Q^{(\pm)} \in {\cal H}_0$ are positive operators corresponding to the
dynamical gauge degrees (they will be specified below). To any $\Omega$
we assign the mapping ${\tilde \Omega}$ in ${\cal S }_k$:
\be
\Psi \ \to \ {\tilde \Omega} \Psi \ =\ \Omega \Omega^o \Psi \ =\
\Omega \Psi \Omega^* \ .
\ee
Explicitely, acting by ${\tilde \Omega}$ and ${\tilde \Omega}^*$ on
$\Psi = fa +ga^* \in {\cal S }_k$ we obtain
\[
{\tilde \Omega} \Psi \ =\ UQ^{(+)} f Q^{(-)} U^* a
\ +\ U Q^{(-)} g Q^{(+)} U^* a^* \ ,
\]
\be
{\tilde \Omega}^* \Psi \ =\ Q^{(+)} U^* f U Q^{(-)} a
\ +\ Q^{(-)} U^* g UQ^{(+)} a^* \ .
\ee

We take the interacting Dirac operator in the form
\be
D \ =\ {\tilde \Omega} D_0 {\tilde \Omega}^* \ .
\ee
Postulating that under gauge transformation
\be
\Omega \ \to \ \omega \Omega \ ,
\ee
we enjoy the transformation property: $D\Psi \to \omega D\Psi \omega^*$.

Formula (40) for the full Dirac operator has the important property that
it preserves the index of the Dirac operator. Any zero-mode
$\Psi_0$ of $D_0$ yields a zero-mode $\Omega^{*-1} \Psi_0$ is  of $D$ and
vice versa. Nonzero-modes appear in pairs, if $\Psi^{(+)}_E$ is an
eigenstate of the full Dirac operator to the eigenvalue $E, E>0 $, then
$\Psi^{(-)}_E = \Gamma \Psi^{(+)}_E$ is an eigenstate to the eigenvalue
$-E$.

The square of the Dirac operator is an even operator in ${\cal S }_k$.
Similarly as in the commutative case, the square of the Dirac operator
can be written as
\be
D^2 \ =\ [\Delta (\Omega ) + q_o F(\Omega )] P_+ \ +\
[\Delta (\Omega ) - q_o F(\Omega )] P_- \ ,
\ee
where $\Delta (\Omega )$ is the noncommutative analog of a Laplace operator
on a sphere with gauge degrees of freedom included, $F(\Omega )$ is the
operator cooresponding to the field strenght and $q_o$ is the coupling
constant. The topological (monopole) contribution to the field strenght is
obtained by putting $\Omega =1$ in $F(\Omega )$. In ${\cal S}_k$ it takes
the same value $q^{-1}_o k$ as in the commutative case.

For a given $J$ and $k$ the spinor field action interacting with gauge
field has the form
\[
S^J_k [\Psi ,\Psi^* ,\Omega ] \ =\
\frac{1}{2J'+1} s{\rm Tr}_N [{\bar \Psi} D \Psi ]
\]
\be
\ +\ \frac{1}{J^2 +J-k^2}
s{\rm Tr}^{J'}_k [ F^2 (\Omega ) P_+ - F^2 (\Omega ) P_- ] \} \ ,
\ee
where $s{\rm Tr}_N$ in the first term denotes the supertrace in the
space of mappings $s{\cal F}_N \to s{\cal F}_N$, and $s{\rm Tr}^J_k$ in the
second one denotes the supertrace in the space of mappings
${\cal S}^{J'}_k \to {\cal S}^{J'}_k$. The factor in the second term
guarantees that the pure topological contribution obtained for $\Omega =1$
is properly normalized to $q_o^{-2} k^2$. The action (46) is gauge invariant.
We can gauge $U$ away and fix the gauge by putting
$\Omega = Q^{(+)} P_+ +Q^{(-)} P_-$.

In the noncommutative case the dynamical gauge degrees of freedom are
described the hermitean operator $\sigma \in {\cal H}_0$ possessing an
expansion like in (35), however now in terms of a noncommutative analogs
of spherical functions: $D^j_{0m} \in {\cal H}_0$, $j=0,1,\dots ,|m|\leq j$,
(see Appendix A). Since, the noncommutative gauge field action
approaches in the commutative limit its commutative form we shall assume
the asymptotic behaviour (36) for the expansion coefficients too.

We define the operators $Q^{(\pm)}$ in the Fock space ${\cal F}$ by their
restrictions $Q^{(\pm)}_{N'}$ to subspaces ${\cal F}_{N'}$. For fixed
$J$ and $k$ in the action appear restrictions with
$N' =J' \pm k \pm \frac{1}{2}$. The maximal value of $N'$ for fixed $J$ and
arbitrary $|k| \le J$ is $2J$. Therefore, it is enough to define the
the operators $Q^{(\pm)}_{N'}$, $N' \leq K$, for some fixed $K>2J$. Namely,
we put
\be
Q^{(\pm)}_{N'} \ =\ \exp \Sigma^{(\pm )}_{KN'} \ , \  N' \leq K \ ,
\ee
where the operators
\be
\Sigma^{(\pm )}_{KN'} \ =\
(\ln \circ {C'}^{KN'}_0 \circ \exp \circ C^{KN'}_0 )(\pm e_o \sigma ) \ ,
\ee
are defined with the help of isometrical mappings
$C^{KN'}_0 :{\cal H}^K_k \to {\cal H}^{N'}_k$ and
${C'}^{KN'}_0 :{\cal H}^{N'j}_k \hookrightarrow {\cal H}^{Kj}_k$ introduced
in Appendix A ($e_o$ is a dimensionless parameter specified later).

This definition guarantees that $\sigma_{KN'} =C^{KN'}_0 \sigma$ posessess
an expansion in ${\cal F}_{N'}$
\be
\sigma_{KN'} \ =\ C^{KN'}_0 \sigma \ =\
\sum_{j=0}^{N'} \sum_{|m|\leq j} b^j_m D^{N'j}_{0m} \ ,
\ee
with the truncated subset of coefficients $\{ b^j_m$, $j = 0,1,\dots ,K$,
$|m|\leq j \}$. Thus, the untruncated modes have the intensities
{\it independent} on $N'$. The field $\sigma$ is for us a primary object,
whereas  $Q^{(\pm)}$ are secondary ones.

\section{Quantization and fermionic determinant}

We quantize the Schwinger model  within the path integral approach. First we
describe the quantization in the commutative case (for details see
\cite{Jay}), and then its noncommutative version.

The quantum expectation values of the field functionals $P[\Xi ]$ (where
$\Xi = \{ \Psi , \Psi^* ,\sigma , \lambda \}$ denotes the collection of all
fields in question) are defined by the formula
\be
\langle P[\Xi ]\rangle \ =\ Z^{-1} \
\sum_k \int (D\Xi )_k \ P[\Xi ] \ e^{-S[\Xi ]} \ ,
\ee
with the summation over all topological winding numbers $\kappa = 2k$ and
the normalization
\be
Z \ =\ {\int (D\Xi )_0 \ e^{-S[\Xi ]}} \ .
\ee
The action $S[\Xi ]$ is given in (43). The symbol
$(D\Xi )_k = (D\Psi D\Psi^* D\sigma D\lambda)_k$ denotes a formal
integration over all field configurations with a given topological number.
The field $\Psi$ can be expanded, e.g. into corresponding eigenfuctions of
the free Dirac operator
\be
\Psi \ =\ \sum^{\infty}_{j=|k|-1/2} \sum_{|m|\leq j} \
(a^{j+}_{km} \Psi^{j+}_{km} \ +\ a^{j-}_{km} \Psi^{j-}_{km} )
\ee
with Grassmannian coefficients $a^{j\pm}_{km}$ (for $j=|k|-1/2$ the
expansion contains only one component corresponding to the zero modes).
Then $D\Psi_k \sim \prod da^{j\pm}_{km}$ means a formal infinite dimensional
Berezin integration. The symbol $D\Psi^*_k$ is defined analogously.

If the field functional $P[\Xi]$ is gauge invariant, then the integrand in
(47) does not depends on a pure gauge prepotential $\lambda$. A gauge
fixing condition in (47) should be imposed, or alternatively, one can
factorize out a pure gauge prepotential $\lambda$, and take in (47) only
the relevant fields $\Xi =\{ \Psi ,\Psi^* ,\sigma \}$. Then
$(D\Xi )_k =(D\Psi D\Psi^*D\sigma )_k$. Expanding $\sigma$ into spherical
fuctions (as in (35)) we introduce a formal infinite dimensional
integration $D\sigma \sim \prod db^j_m$ (with $b^j_{-m} =(-1)^m b^j_m$).

Thus, the expectation values are defined only formally, and some
regularization procedure is needed. In \cite{Jay} were calculated in this
way various non-perturbative quantities. If the functional $P[\Xi ]$ in
question is a polynomial in spinor fields then the calculation over
$(D\Psi D\Psi^* )_k$ can be performed explicitely. This leads to the
fermionic determinants and effective actions which are needed for the
calculation of fermionic condensates.

The gauge field effective action $S_k^{reg}[\sigma ]$ is obtained
by integration over $D\Psi D\Psi^* $ in (47) and is given as
\be
S_k^{eff} [\sigma ] \ =\ S_k [\sigma ] \ +\ \Sigma_k^{reg} [\sigma ]\ ,
\ee
where $S_k [\sigma ]$ is the classical gauge field action and the second
term represents the quantum correction	defined by the equation
\be
C^{reg}_k {\rm det}_k D \ =\ \exp (-\Sigma_k^{reg} [\sigma ]) \ .
\ee
Here $C^{reg}_k$ is a field independent regularizating factor and
${\rm det}_k D$ denotes the product of non-zero eigenvalues of $D$ in the
sector with topological winding number $\kappa = 2k$. The formula derived
in \cite{Jay} reads
\be
\Sigma_k^{reg} [\sigma ] \ =\ 2 \int d\nu (\nabla \sigma )^2 -
\pi i |\kappa | - \frac{1}{2}| \kappa |^2 - 2 |\kappa | \ln (|\kappa |!)
+ 2 \sum_{n=1}^{|\kappa |} n \ln n \ .
\ee

The effective action appears in formulas for many important quantities, e.g.
the fermionic condensate - the mean value of the field functional
$(\Psi, \Psi )=\int d\mu {\bar \Psi} \Psi$. The simplest way how to
calculate it consists in adding a "mass" term $m(\Psi, \Psi )$ to the
action and using the formula
\be
\langle (\Psi, \Psi ) \rangle \ =\ -\frac{\partial}{\partial m}
\langle e^{-m(\Psi, \Psi )} \rangle {\big|}_{m =0} \ .
\ee
There are two equal contributions $\langle (\Psi, \Psi )\rangle_{\pm 1/2}$
appearing in the numerator in (47) for $k=\pm 1/2$.

In the noncommutative case we shall start from the general formula
(47), however there are important differences. We fix $J$ and restrict the
admissible range of $k$ to $|k| \leq J$. The symbol
$(D\Xi )^J_k = (D\Psi D\Psi^* D\sigma)^J_k$ denotes an integration over
all field configurations with a given topological number:

(i) The spinor fields $\Psi \in {\cal S}^{J'}_k$ and
$\Psi^* \in {\cal S}^{J'}_{-k}$ can be expanded into corresponding
eigenfuctions of the free Dirac operator with independent Grassmann
coefficients $a^{\pm j}_m$, $a^{*\pm j}_m$. The fermionic part of the
integration measure $(D\Xi )^J_k$ denotes the finite Berezin integral
\be
(D\Psi D\Psi^* )^J_k \ =\ \prod_{|m|\leq |k|-1/2} \frac{da^0_m}{\sqrt{J}}
\prod_{j=|k|+1/2}^{J'} \prod_{|m|\leq j}
\frac{da^{\pm j}_m}{\sqrt{J}} \frac{da^{*\pm j}_m}{\sqrt{J}} \ ,
\ee
(the first product comes from the contributions of over zero-modes and is
present only for $k\neq 0$).

(ii) We restrict $\sigma$ to ${\cal F}_K$ with some fixed $K>2J$. Expanding
$\sigma$ as in (46) with expansion coefficients $b^j_m =(-1)^m b^j_{-m}$ we
put
\be
(D\sigma )^J_k \ =\ \prod_{j=0}^K db^j_0 \prod_{m=1}^j db^j_m \ .
\ee

There are no problems with the gauge fixing. The gauge group is isomorphic
to $SU(K+1)$ possessing a finite volume which can be factorized out. The
number of all modes, the dimension of the measure $(D\Xi )^J_k$, is finite,
and consequently there are no ultraviolet divergencies. This allows to
calculate straightforwardly various non-perturbative quantities.

Let the observable $P[\Xi ]$ in (47) be independent of fermion fields. We
perform a bosonization of the model by integrating out the fermionic
degrees of freedom and obtaining an effective action
$S^{eff}_{Jk} [\sigma ] = S_k^J [\sigma ] + \Sigma_k^J [\sigma ]$. Here
$S_k^J [\sigma ]$ is the classical noncommutative gauge action (the second
term in (43)) and $\Sigma_k^J [\sigma ]$ is the quantum correction given as
\be
\exp (-\Sigma_k^J [\sigma ])\ =\ \frac{{\rm det}_k D}{{\rm det} D_0 } \ =\
\frac{\int (D'\Psi D'\Psi^* )^J_k
e^{-({\bar \Psi},D\Psi )^J_k}}{\int (D\Psi D\Psi^* )^J_0
e^{-({\bar \Psi},D_0 \Psi )^J_0}} \ ,
\ee
where $(D'\Psi D'\Psi^* )^J_k$ denotes the integration over non-zero modes.
It is more convenient to take instead of
$D={\tilde \Omega} D_0 {\tilde \Omega}^*$ the shifted Dirac operator
${\tilde \Omega} (D_0 +m ){\tilde \Omega}^*$, possessing $2|k|$ modes with
$E=m$ instead of zero-modes. Then
\be
\exp (-\Sigma_k^J [\sigma ])\ =\ \lim_{m\to 0} \frac{1}{m^{2|k|}}
\frac{\int (D\Psi D\Psi^* )^J_k
e^{-({\bar \Psi},{\tilde \Omega}(D_0 +m){\tilde \Omega}^*
\Psi )^J_k}}{\int (D\Psi D\Psi^* )^J_0 e^{-({\bar \Psi},D_0 \Psi )^J_0}} \ .
\ee
We rewritte it as a product of two factors. The first one
\be
\lim_{m \to 0} \frac{\int (D\Psi D\Psi^* )^J_k
e^{-({\bar \Psi},{\tilde \Omega}(D_0 +m){\tilde \Omega}\Psi )^J_k}}
{\int (D\Psi D\Psi^* )^J_k e^{-({\bar \Psi},(D_0 +m)\Psi )^J_k}} \ =\
{\rm det}{\tilde \Omega}\ {\rm det}{\tilde \Omega}^*
\ee
do not depend on $m$.

In Appendix B we derived the formula for the rescaling mapping
\be
{C'}^{KN'}_0 \ =\ \exp
\{ -\sum^\infty_{k=0} [(N'+1)^{-2k-1} - (K+1)^{-2k-1} ]S_k (\Delta )\} \ ,
\ee
with a known polynomials $S_k (\Delta )$ in Laplace operator
$\Delta =X_i^2$, and an analogous formula for $C^{KN'}_0$ differing just by
the sign in the exponent. Using them we calculated in Appendix B the
commutative limit $J \to \infty$ of the first factor. Taking
$K = O(J^\varepsilon )$, $\varepsilon >1$, and normalizing the constant
$e_o$ properly by $e^2_o =q^2_o (2J+1)^{-1}$ we obtained the leading term
in the asymptotic form of (58):
\be
{\rm det}{\tilde \Omega}\ {\rm det}{\tilde \Omega}^* \ =\
\exp \{ -2q_o^2 \int d\nu (X_i \sigma )^2 \ +\ o(J^{-1} ) \} \ .
\ee
The second factor gives with $\kappa = 2k$
\[
\lim_{m \to 0} \frac{1}{m^{2|k|}}
\frac{\int (D\Psi D\Psi^* )^J_k e^{-({\bar \Psi},(D_0 +m)\Psi )^J_k}}{\int
(D\Psi D\Psi^* )^J_0 e^{-({\bar \Psi},D_0 \Psi )^J_0}}
\ =\ J^{2|k|^2} \ \frac{\det_k D_0}{\det_0 D_0} \ =\
\]
\be
\exp [\pi i |\kappa | +\frac{1}{2}| \kappa |^2 + 2 |\kappa |
\ln (|\kappa |!) - 2 \sum_{n=1}^{|\kappa |} n \ln n \ +\ o(J^{-1}) ]\ .
\ee
Eqs. (60) and (61) yields in the limit $J \to \infty$ the well-known
expression for the effective action (see e.g. \cite{Jay}).

\section{Concluding remarks}

In this article we described a noncommutative regularization of the
Schwinger model on a noncommutative sphere. We constructed Connes real
spectral triple $({\cal A}, {\cal H}, D_0)$ supplemented by the the grading
$\Gamma$ and the antilinear isometry ${\cal J}$ (identified with the
chiralitity operator and charge conjugation). An important new aspect with
respect to our previous work \cite{GKP1}-\cite{GKP3} lies in the fact that as
the spectral algebra ${\cal A}$ we take an infinitedimensional associative
algebra - the even part of the superalgebra ${\cal B}={\cal U}(u(2|1))$.
We have worked within its canonical realization.

The elements $A B^o \in {\cal A} \otimes {\cal A}^o$ act in the Hilbert
space ${\cal H}$ of spinor fields as the left multiplications by $A$ and
the right one by $B^*$. Introducing the differentials $dA =[D_0 ,A]$ and
$dB^o =[D_0 ,B^o ]$  we obtain that in our case the condition
$[A,dB^o ]=[dA,B^o ]=0$, usually required in the noncommutative geometry,
is violated.

We gauged the model in a standard way: $\Psi \to \omega \Psi \omega^*$,
with $\omega$ unitary. The compensating gauge fields $A$ enter the full
Dirac operator which can be expressed as
$D=D_\alpha {\bar D}_\alpha + D^*_\alpha {\bar D}^*_\alpha$ with
$D_\alpha =V_\alpha + q_o A_\alpha$ and
${\bar D}_\alpha ={\bar V}_\alpha + q_o {\bar A}_\alpha$. Such a form of
$D$ with general $A_\alpha$ and ${\bar A}_\alpha$ was recently proposed in
\cite{K} within supersymmetric extension of the Schwinger model. This
formulation contains besides physical fields auxiliary fields. We eliminated
them by expressing $A_\alpha =UQ[V_\alpha ,U^{-1} Q^{-1} U^{-1} ]$ and
${\bar A}_\alpha =UQ^{-1} [{\bar V}_\alpha ,Q U^{-1} ]$ in terms of
prepotentials. Moreover, our choice preserves the index of the Dirac operator.

The physical model is determined if we choose the action for the fields in
question. The evaluation of the action we performed by restricting the
fields to a particular finite dimensional mappings in the Fock space. This
required sequences of linear $SU(2)$-invariant embeddings of various
representations of ${\cal A}$. We evaluated the rescaling coefficients of
these embeddings which are necessary for the evaluation of the action.

The action of the resulting noncommutative Schwinger model contains finite
number of modes for each field in question. This allows nonperturbatively
to quantize the model. As an application we calculated the chiral anomaly
and the effective actions. Although we obtained standard results in the
commutative limit our interpretation is very different:

- in the commutative case the nontrivial value of the fermionic
determinants is due to the singularity of the operator
$\exp (-\varepsilon D^2 )$ for $\varepsilon \to 0$, and this leads to the
chiral symmetry breaking,

- in our approach the chiral anomaly follows straightforwardly from the
basic postulates - its appearance is a direct consequence of the fact
that the isometric embeddings and exp-log mappings do not commute.

The noncommutative formulation of the Schwinger model leads not only to
the ex\-pec\-ted ultraviolet regularization, but straightforwardly to
high\-ly nonperturbative results. It provides a better understanding of
the procedure of regularization and renormalization, and the mechanisms of
bosonization and chiral symmetry breaking. It would be interesting to
extend our investigations in various directions, e.g. to include gravity
along lines presented in \cite{Con3}. \\

\noindent {\bf Appendix A}

\noindent Here we shall describe various $SU(2)$ representations in the
space ${\cal H}_{k}$. Any operator $\Phi$ from this space can be expanded as
$\Phi =\sum b^j_m D^j_{km}$ with $m = -j, -j+1, \dots ,+j$,
$j = |k|, |k|+1, \dots$. The operators $D^j_{km}$ can be constructed as
follows:

(i) The operator
\be
D^j_{k,-j} \ =\ [(2j+1)!/(j+k)!(j-k)!]^{1/2}\,
{\hat \chi}^{*j+k}_2 {\hat \chi}^{j-k}_1
\ee
is the lowest weight with respect to the adjoint action of the operator
$X_0$ because $X_0 D^j_{k,-j} =[x_0 ,D^j_{k,-j} ]= -jD^j_{k,-j}$ and
$X_- D^j_{k,-j} =[x_- ,D^j_{k,-j} ]= 0$ (here
${\hat \chi}_\alpha = \chi_\alpha (\chi+ \chi )^{-1/2}$ and
${\hat \chi}^*_\alpha = (\chi+ \chi )^{-1/2} \chi^*_\alpha$).

(ii) For a given $j$ all other $D^j_{km}$ can be obtained by a repeated
action of $X_+$,
\be
D^j_{km} \ =\ [(j+m)!/(j-m)!(2j)!]^{1/2} \ X^{j-m}_+ D^j_{k,-j} \ .
\ee
The operators $D^j_{km}$, $|m| \leq j$ (for a given $j$ and $k$) span a
representation space $V_k^j$ of the $SU(2)$ representation with spin $j$.

By definition we take $\{ D^j_{km}, j = |k|, |k+1, \dots ,|m|\leq j \}$
as an orthonormal basis in ${\cal H}_k$. Thus,
\be
{\cal H}_k \ =\ \bigoplus^{\infty}_{j=|k|} V_k^j \ .
\ee

Restricting the action of superfields $\Phi \in {\cal H}_k$ from
$s{\cal F}$ to $s{\cal F}_N$ we obtain a space of mappings
$s{\cal F}_N \to s{\cal F}_M$, $M=N+2k$. We shall denote it by
${\cal H}^J_k$, $k=\frac{1}{2}(M-N)$, $J=\frac{1}{2}(M+N)$. The expansion
(64) is now truncated:
\be
{\cal H}^J_ k \ =\ \bigoplus^J_{j=|k|} V^{Jj}_k \ ,
\ee
where $V^{Jj}_k$ denotes the space spanned by the operators (63)
restricted to $s{\cal F}_N$. If we introduce in ${\cal H}^J_k$ the scalar
product
\be
(\Phi_1 ,\Phi_2 )^J_k \ =\ \frac{1}{J+1} {\rm Tr} (\Phi^*_1 \Phi_2 )\ ,\
\ee
then the orthonormal basis in $V^{Jj}_k$ is given as
\be
D^{Jj}_{km} \ =\ c^J_{kj} D^j_{km} \ ,\ m = -j, \dots ,+j \ ,
\ee
with
\be
c^J_{kj} \ =\ [(J+1)(J+k)!(J-k)!/(J+j+1)!(J-j)!]^{1/2} \ .
\ee
Summing up over $j=|k|, \dots ,J$ one obtains aan orthonormal basis in
${\cal H}^J_k$.

Taking $K \ge J$ we define the isometrical projection operator
$C^{KJ}_k :{\cal H}^K_k \to {\cal H}^J_k$
by $D^{Kj}_{km} =C^{KJ}_k D^{Jj}_{km}$, and by
${C'}^{KJ}_k :{\cal H}^{Jj}_k \to {\cal H}^{Kj}_k$ we denote the reversed
isometrical embedding. The operators $C^{KJ}_k$ and ${C'}^{KJ}_k$ are both
diagonal with the eigenvalues $c^K_{kj}/c^J_{kj}$ and $c^J_{kj}/c^K_{kj}$
respectively, as follows from the equation
\be
c^J_{kj} D^{Kj}_{km} \ =\ c^K_{kj} D^{Jj}_{km} \ ,\ j \le J \le K \ .
\ee

Now we determine the spectrum of the free Dirac operator. Since,
$[X_0 , D_0 ]$ $=0$, we classify the eigenstates of $D_0$ according to the
eigenvalues values of $X_0$. The spinor field
\be
\Psi \ =\ \alpha \chi^{*m}_1 \chi^{n-1}_2 a \
+\ \beta \chi^{*m-1}_1 \chi^n_2 a^* \
\ee
is the lowest weight of $X_0$ because, $X_0 \Psi =[x_0,\Psi ]=(m+n-1)\Psi$
and $X_- \Psi =[x_-,\Psi ]=0$. Moreover,
\be
D_0 \Psi \ =\ \alpha m\chi^{*m-1}_1 \chi^n_2 a^* \
+\ \beta n\chi^{*m}_1 \chi^{n-1}_2 a \ .
\ee
This is an eigenstate to the value $E =\pm \sqrt{mn}$ provided that
$\alpha \sqrt{m} = \beta \sqrt{n}$. The spinor field lowest weight
eigenstates are
\be
\Psi^{j\pm}_{k,-j} \ =\ c(\sqrt{n} \chi^{*m}_1 \chi^{n-1}_2 a \
\pm \ \sqrt{m} \chi^{*m-1}_1 \chi^n_2 a^* )\ ,
\ee
where $j=\frac{1}{2}(m+n-1)$, $k=\frac{1}{2}(m-n)$ (if $k\neq 0$ we obtain
for $m=0$ or $n=0$ zero modes corresponding to the first or second term in
(74) respectively). The remaining eigenstates of $D_0$ to the same
eigenvalue are given by the standard formula
\be
\Psi^{j\pm}_{km} \ =\ \left[ (j+m)!/(j-m)!(2j)! \right]^{1/2} \
X^{j-m}_+ \Psi^{j\pm}_{k,-j} \ .
\ee
The admissible values of $j$ then are $j = |k|, |k|+1,\dots J$. The
normalization constant $c$ in (72) is specified if we restrict the
space of spinor fields ${\cal S }_k$ to the space ${\cal S }^J_k$ with
the inner product:
\be
(\Psi_1 ,\Psi_2 )^J_k \ =\
\frac{1}{2J+1}\, s{\rm Tr}({\bar \Psi}_1 \Psi_2 ) \ .\\
\ee

\noindent {\bf Appendix B}

\noindent From the formula (68) for the rescaling coefficitients
it follows that
\[
c^N_{0j} \ =\ \exp \left[ \frac{1}{2} \sum_{n=1}^j
\ln \frac{1-n/(N+1)}{1+n/(N+1)}  \right] \ =\
\]
\be
\exp \left[ -\sum_{k=0}^\infty (N+1)^{-2k-1} S_k (j(j+1)) \right] \ ,
\ee
where $S_k (t) = [(2k+1)(2k+2)]^{-1} t^k + \dots$ are polynomials defined
by
\[
S_k (j(j+1)) = \frac{1}{2k+1} \sum_{n=1}^j n^{2k+1} \ .
\]
The eigenvalues of the operator ${C'}^{KN'}_0$ can be expressed as
\be
c^{N'}_{0j}/c^K_{0j} \ =\ \exp
\{ -\sum^\infty_{k=0} [(N'+1)^{-2k-1} - (K+1)^{-2k-1} ]S_k (j(j+1))\} \ .
\ee
The expansion in (75) is convergent provided $j \le N' \le K$.
The eigenvalues of the operator $C^{KN'}_0$ differs just by the sign in
front of the sum in exponent.

Replacing $j(j+1)$ by the (noncommutative) Laplace operator on a sphere
$\Delta = X^2_i$ one obtains directly from (76) the formulas for the operators
${C'}^{KN}_0$ and $C^{KN'}_0$. This allows to derive the leading term
formula
\[
Q^{(\pm )}_{N'} \ =\
{C'}^{KN'}_0 \exp \{ C^{KN'}_0 (\pm e_o \sigma ) \} \ =\
\]
\[
[1 - \frac{K-N'}{2(K+1)(N'+1)}\Delta +\dots ]
\exp [\pm e_o (1 + \frac{K-N'}{2(K+1)(N'+1)}\Delta +\dots )\sigma ]
\]
\be
\ =\ \exp [\pm e_o \sigma  -e^2_o \frac{K-N'}{2(K+1)(N'+1)}
(X_i \sigma )^2 + o(K^{-2}, {N'}^{-2} )] \ .
\ee

Using the well known formula
$\det A\otimes B = (\det A)^M (\det B)^N$ (valid for matrices
$A \in {\rm Mat}(N\otimes N)$ and $B \in {\rm Mat}(M\otimes M)$) and the
definition of the operators
${\tilde \Omega}:{\cal S}^{J'}_k \to {\cal S}^{J'}_k$ it can be easily seen
that in the gauge $U=1$ we have 
\[
\det {\tilde \Omega} \ =\ \det [\Omega \Omega^o ] \ =\
\]
\be
({\det}_{M-1} Q^{(+)} )^{N+1} ({\det}_N Q^{(-)} )^M
({\det}_M Q^{(-)} )^N ({\det}_{N-1} Q^{(+)} )^{M+1} \ ,
\ee
where $N=J'-k+\frac{1}{2}$ and $M=J'+k+\frac{1}{2}$ as usually. Inserting
here (77) we obtain \\
\[
\ln \det {\tilde \Omega} \ =\ e_o [(N+1){\rm Tr}_{M-1} \sigma -
M{\rm Tr}_N \sigma -N{\rm Tr}_M \sigma (M+1){\rm Tr}_{N-1} \sigma ]
\]
\[
- \frac{e^2_o}{2(K+1)} \left[ \frac{(N+1)(K-M+1)}{M} {\rm Tr}_{M-1}
(\sigma )^2 +\frac{M(K-N)}{N+1} {\rm Tr}_N (\sigma )^2 \right. +
\]
\be
\left. \frac{N(K-M)}{M+1} {\rm Tr}_M (\sigma )^2 +
\frac{(M+1)(K-N+1)}{N} {\rm Tr}_{N-1} (\sigma )^2 + \right]
\ee
Taking $K = O(J^\varepsilon )$, $\varepsilon >1$, normalizing the constant
$e_o$ by $e^2_o =q^2_o (2J+1)^{-1}$ and using relations
\[
\frac{1}{N'} {\rm Tr}_{N'} \sigma \ \to \ \int d\nu \sigma \ \ ,\ \
\frac{1}{N'} {\rm Tr}_{N'} (\nabla \sigma )^2 \ \to \
\int d\nu (\nabla \sigma )^2 \ ,\
\]
(which are valid for $N' \to \infty$ due to the asymptotic behaviour (36))
we obtain in the limit $J\to \infty$ the asymptotic formula
\be
\ln \det {\tilde \Omega} \ =\ -q^2_o \int d\nu \
(X_i \sigma )^2 \ +\ o(J^{-1} ) \ ,
\ee
leading sraightforwardly to eq. (60).

For a given $J$ and $k\neq 0$ the free Dirac operator possesses non-zero
modes $E^{J\pm}_{jk} =\pm \sqrt{(j+\frac{1}{2})^2-k^2}$,
$j=|k|+\frac{1}{2},|k|+\frac{3}{2},\dots ,J$,  with the multiplicity
$2j+1$. Putting $j' =j+\frac{1}{2}$ and $J' =J+\frac{1}{2}$ we obtain
\[
J^{2|k|^2} \ \frac{\det_k D_0}{\det_0 D_0} \ =
(-1)^{2k} J^{2|k|^2} \
\frac{\prod_{j'=|k|+1}^{J'}
(j^2 -|k|^2)^{2j'}}{\prod_{j'=1}^{J'} {j'}^{4j'}} \ =
\]
\be
(-1)^{2k}\prod_{m=1}^{|k|}
\left( \frac{1+m/J'}{1+(m-|k|)/J'} \right)^{2J+2m}
\ \prod_{n=1}^{2|k|} n^{2|k|-2n} \
\frac{J^{2|k|^2}}{\prod_{n=1}^{|k|} (J'+n)^{2|k|}} \ .
\ee
Denoting $\kappa =2k$ we obtain in the limit $J\to \infty$ desired eq. (61).


\begin{thebibliography}{99}
\bibitem{Con1} A. Connes, Publ. IHES {\em 62} (1986) 257.
\bibitem{Con2} A. Connes, Geometrie Noncommutative (Inter Editions, Paris
1990).
\bibitem{Con3} A. Connes, {\it Gravity coupled with matter and the
foundation of non commutative geometry}, hep-th/9603053.
\bibitem{DV} M. Dubois-Violette, C. R. Acad. Sci. Paris {\em 307}, Ser. I
(1988) 403.
\bibitem{DKM} M. Dubois-Violette, R. Kerner and J. Madore, Journ. Math. Phys.
{\em 31} (1990) 316.
\bibitem{Ber} F. A. Berezin, Commun. Math. Phys. {\em 40} (1975) 153.
\bibitem{Hoppe} J. Hoppe, Elem. Part. Res. J. {\em 80} (1989) 145 .
\bibitem{Mad} J. Madore, Journ. Math. Phys. {\em 32} (1991) 332.
\bibitem{GP1} H. Grosse and P. Pre\v{s}najder, Lett. Math. Phys.
{\em 28} (1993) 239.
\bibitem{GKP1} H. Grosse, C. Klim\v{c}\'{\i}k  and P. Pre\v{s}najder,
Int. Journ. Theor. Phys. {\em 35} (1996) 231.
\bibitem{GKP2} H. Grosse, C. Klim\v{c}\'{\i}k  and P. Pre\v{s}najder,
Commun. Math. Phys. {\em 178} (1996) 507.
\bibitem{GKP3} H. Grosse, C. Klim\v{c}\'{\i}k  and P. Pre\v{s}najder,
Commun. Math. Phys. {\em 185} (1997) 155.
\bibitem{GP2} H. Grosse and P. Pre\v{s}najder, Lett. Math. Phys.
{\em 33} (1995) 171.
\bibitem{Jay} C. Jayewardena, Helv. Phys. Acta {\em 61} (1988) 638.
\bibitem{GM} H. Grosse and J. Madore, Phys. Lett. {\em B283} (1992) 218.
\bibitem{GKP4} H. Grosse, C. Klim\v{c}\'{\i}k  and P. Pre\v{s}najder,
{\it Finite gauge model on truncated sphere}, Proc. of Schladming School,
p. 279, Lect. Notes in Phys. (Springer-Verlag Berlin 1996).
\bibitem{K} C. Klim\v{c}\'{\i}k, {\it Gauge theories on the noncommutative
sphere}, preprint IHES/P/97/77, hep-th/9710153.
\bibitem{Haw} E. Hawkins, {\it Quantization of equivariant vector bundles},
preprint Pensylvania State Univ. CGPG 97-98/8-1, q-alg/9708030.
\bibitem{Ber2} F. A. Berezin,  Introduction to Superanalysis, (Reidel,
Dordrecht 1987).
\bibitem{Rit} M. Scheunert, W. Nahm and V. Rittenberg, Journ. Math. Phys.
{\em 18} (1977) 156.
\end{thebibliography}
\end{document}